\documentclass{iopconfser}

\newcommand\be{\begin{equation}}
\newcommand\ee{\end{equation}}
\newcommand{\bea}{\begin{eqnarray}}
\newcommand{\eea}{\end{eqnarray}}

\newcommand{\nn}{\nonumber}
\newcommand{\pd}{\partial}

\usepackage{cite}
\usepackage{graphicx}

\begin{document}

\title{Multifield Cosmology and the Dark Universe}

\author{Lilia Anguelova$^{1}$}

\affil{$^1$Institute for Nuclear Research and Nuclear Energy,
Bulgarian Academy of Sciences, Sofia 1784, Bulgaria}

\email{anguelova@inrne.bas.bg}

\begin{abstract}
Multifield models, arising from multiple scalars interacting with gravity, provide a rich theoretical framework for addressing fundamental problems in modern cosmology. A key role in this regard is played by the so called rapid turn regime, which is characterized by background solutions with strongly non-geodesic field-space trajectories. We review the implications of this regime for a number of problems relevant for cosmological inflation, dark matter and dark energy. We focus, in more detail, on a class of exact rapid-turn solutions that give a model of dynamical dark energy. In this model, the sound speed of the dark energy perturbations is reduced compared to the speed of light, which leads to observational differences from a cosmological constant even for an equation-of-state parameter very close to $-1$\,. Furthermore, this model holds promise for the simultaneous alleviation of two prominent cosmological tensions.
\end{abstract}

\section{Introduction}

Modern cosmology is blessed with a huge amount of accurate observational data. This has enabled a very precise determination of the values of the cosmological parameters, which characterize the composition and evolution of the Universe. The resulting standard cosmological model, called $\Lambda$CDM, provides a good phenomenological description, overall. However, it also raises deep questions for fundamental physics. For example, understanding the nature of dark energy and dark matter (the dominant components in the energy density of the Universe) is still an open problem, which could have profound implications for our theories of gravity and elementary particles. Furthermore, in recent years the increasing precision of the observational data has led to a number of so called cosmological tensions. A tension, in this context, means a discrepancy between the values of the same cosmological parameter inferred from different data sets, while assuming the standard $\Lambda$CDM model. All of these issues motivate the study of alternative cosmological models.

The framework of multifield cosmology provides a particularly rich class of such models. They arise from the interaction of gravity with many scalar fields. Models with a single scalar coupled to gravity (so called `single-field models') have long been studied for the purposes of describing cosmological inflation or dynamical dark energy. However, recent theoretical developments have led to the realization that multifield models (i.e., models containing more than one scalar) can lead to novel phenomenological effects, while also being preferred from the perspective of fundamental physics \cite{GK,OPSV,AP,BPR}. The main difference from the single-field case is that in the multifield context one can have solutions of the background equations of motion, whose field-space trajectories are strongly non-geodesic. Such trajectories exhibit rapid turning in field space, which can lead to slow-roll inflationary solutions even when the scalar potential does not have flat directions \cite{AP,HP}. Furthermore, inflationary field-space trajectories with sharp turns can induce primordial black hole generation in the early Universe \cite{PSZ,FRPRW,LA,LA2}. Also, interestingly, multifield models of dynamical dark energy can have observational features that distinguish them from a cosmological constant, even for an equation-of-state parameter very close to $-1$; see \cite{ASSV}.

\eject
We outline how the above new effects occur in two-field cosmological models with rapid-turning solutions, focusing more extensively on the dynamical dark energy model of \cite{ADGW,ADGW2}. We review the class of exact background solutions this model relies on. The key to obtaining the solutions analytically is the use of a hidden symmetry found in \cite{ABL1}. In addition, we discuss the perturbations of the dark energy scalars around these backgrounds. In \cite{ADGW2} it was shown that the sound speed of these perturbations is reduced compared to the speed of light, everywhere in the parameter space of the solutions. This implies the existence of a certain dark energy clustering scale (of order Gpc) within the observable Universe. Moreover, as argued in \cite{ADGW2}, this class of dynamical models is very promising for the simultaneous alleviation of the Hubble and $\sigma_8$ cosmological tensions. That makes it very appealing and worthy of further investigation. In concluding remarks, we discuss new observational challenges and directions for future studies.

\section{Multifield cosmological models}

We consider a class of cosmological models, which arises from Einstein gravity minimally coupled to multiple scalar fields. The key distinction from the single-scalar models is that one can have a curved field-space metric. The standard action governing this system is:
\be \label{Action_gen}
S = \int d^4x \sqrt{-\det g} \left[ \frac{R}{2} - \frac{1}{2} G_{IJ} \pd_{\mu} \phi^I \pd^{\mu} \phi^J - V (\{ \phi^I \}) \right] \,\,\, ,
\ee
where $g_{\mu \nu}$ is the spacetime metric with indices $\mu,\nu = 0,1,2,3$ , while $G_{IJ}$ is the metric on the manifold parametrized by $n$ scalars $\phi^I$ with indices $I,J = 1,...,n$. As usual, we assume that the background spacetime metric and background scalars have the form:
\be \label{metric_g}
ds^2_g = -dt^2 + a^2(t) d\vec{x}^2 \qquad , \qquad \phi^I = \phi^I_0 (t) \quad 
\ee 
with $a(t)$ being the scale factor. We also recall the standard definition of the Hubble parameter: $H = \dot{a} / a$\,.

The background field equations, that follow from (\ref{Action_gen})-(\ref{metric_g}), are:
\be \label{EoMs}
D_t \dot{\phi}^I_0 + 3 H \dot{\phi}_0^I + G^{IJ} V_J = 0 
\ee
and 
\be \label{EinstEqs}
G_{IJ} \dot{\phi}^I_0 \dot{\phi}^J_0 = - 2 \dot{H}  \qquad  ,  \qquad  3 H^2 + \dot{H} = V \,\,\, ,
\ee
where we have used the notation $\dot{} \equiv \pd_t$\,, $V_J \equiv \pd_J V$\,, $\pd_J \equiv \pd_{\phi_0^J}$\,. Also, the definition of $D_t A^I$, for any vector $A^I$ in field space, is:
\be
D_t A^I \equiv \dot{\phi}_0^J \,\nabla_J A^I = \dot{A}^I + \Gamma^I_{JK} \dot{\phi}_0^J A^K \,\,\, ,
\ee
where $\Gamma^I_{JK}$ are the Christoffel symbols of the field-space metric $G_{IJ}$.

Important physical properties of the space-time solutions are determined by the shape of their field-space trajectories. To illustrate this in the two-field case, which we will focus on in the following sections, let us introduce a convenient orthonormal basis in field space, determined by the tangent and normal vectors to the field-space trajectory $\left( \phi^1_0 (t),\phi^2_0 (t) \right)$:
\be \label{TN_def}
T^I = \frac{\dot{\phi}^I_0}{\dot{\sigma}} \qquad {\rm and} \qquad N_I = (\det G)^{1/2} \epsilon_{IJ} T^J \quad ,
\ee
where $\dot{\sigma} \equiv \sqrt{G_{IJ} \dot{\phi}^I_0 \dot{\phi}^J_0}$ and $I,J = 1,2$\,. Then, the turning rate of the trajectory can be defined as:
\be
\Omega = - N_I D_t T^I \quad .
\ee
Note that $\Omega$ measures the deviation from a geodesic in field space. Another important set of parameters, characterizing the background solutions, is given by the following quantities:
\be \label{SR_RT_par}
\varepsilon \equiv -\frac{\dot{H}}{H^2} \qquad , \qquad \eta_{\parallel} \equiv - \frac{\ddot{\sigma}}{H \dot{\sigma}} \qquad , \qquad \eta_{\perp} \equiv \frac{\Omega}{H} \quad .
\ee
The first two, $\varepsilon$ and $\eta_{\parallel}$\,, are the standard slow-roll parameters in single-field inflation with inflaton $\sigma (t)$. The third parameter, the dimensionless turning rate $\eta_{\perp}$\,, brings in qualitatively new features when $\eta_{\perp}^2 \gg 1$\,. This is known as the rapid turn regime. It occurs when background solutions have strongly non-geodesic field-space trajectories.

\section{Novel features of rapid turn models}

Multifield cosmological models have been studied in the literature for a long time. However, it is a rather recent realization that non-geodesic field space trajectories can lead to novel effects that are of great phenomenological interest. In this section we review three main examples of such new features, which occur due to rapid-turning trajectories in two-field models. 

\subsection{Slow roll on steep potentials}
A key feature of the rapid turn regime is that flat directions in the scalar potential are no longer necessary for slow-roll spacetime expansion. Namely, during a slow-roll and rapid-turn regime, i.e. when $\varepsilon\,,\,|\eta_{\parallel}| \ll 1$ and $\eta_{\perp}^2 \gg 1$\,, one has \cite{HP}:
\be \label{Ep_EpV}
\varepsilon \,\approx \,\varepsilon_V \left( 1 + \frac{\eta^2_{\perp}}{9} \right)^{-1} \quad , \quad {\rm where} \quad \varepsilon_V = \frac{1}{2} \frac{G^{IJ}V_I V_J}{V^2} \,\,\,\, .
\ee 
This means that $\varepsilon$ can be small, even when $\varepsilon_V$ is of order $1$ or larger, as long as $\eta_{\perp}^2$ is large. In other words, rapid-turning trajectories allow us to realize cosmological inflation even in models with (very) steep potentials. This could facilitate the embedding of effective models of slow-roll inflation in fundamental theoretical set-ups, like string compactifications. Furthermore, as pointed out in \cite{AP}, relation (\ref{Ep_EpV}) shows that, in the multifield case, slow-roll cosmic acceleration can be compatible with the swampland conjectures \cite{GK,OPSV}, which require $\varepsilon_V \sim {\cal O} (1)$\,.\footnote{Note, however, that in string compactifications, in which rapid-turn cosmic acceleration is realized near the boundary of moduli space, satisfying the swampland distance conjecture may be problematic \cite{FLNS}.} For this reason, the considerations of \cite{AP} (see also \cite{BPR}) sparked a lot of interest in rapid turn inflation.

\subsection{Primordial black hole generation}

Another interesting phenomenon, that rapid turning in field space can lead to, is the generation of primordial black holes (PBHs), which are a natural and promising candidate for dark matter.\footnote{For other interesting possibilities for multi-component dark matter, as well as Beyond Stadard Model (BSM) physics relevant for cosmology, see the review \cite{MK}.} PBHs can form from large fluctuations during cosmological inlfation. To induce PBH-generation, those fluctuations have to be about $10^7$ times larger than the perturbations in the cosmic microwave background (CMB) radiation. In single-field inflationary models, such an enhancement of the amplitude of the perturbations is a challenge. On the other hand, in the multifield case, it is easily achieved when the background solution has a field-space trajectory that exhibits a brief sharp turn \cite{PSZ,FRPRW,LA,LA2}.

To outline this mechanism, let us decompose the inflatons into a background and perturbations:
\be \label{Pert_def}
\phi^I (t, \vec{x}) = \phi^I_0 (t) + \delta \phi^I (t, \vec{x}) \,\,\, .
\ee
In the basis (\ref{TN_def}), the fluctuations have the following components:
\be \label{Pert_TNproj}
\delta \phi_{\parallel} = T_I \delta \phi^I \qquad , \qquad \delta \phi_{\perp} = N_I \delta \phi^I \quad ,
\ee
called respectively adiabatic and isocurvature perturbations. The metric fluctuations can be written, in comoving gauge, as:
\be
g_{ij} (t, \vec{x}) = a^2(t) \left[ ( 1 + 2 \zeta ) \delta_{ij} + h_{ij} \right] \,\,\, ,
\ee
where $i,j = 1,2,3$ are the spatial indices, $\zeta$ is the curvature perturbation and $h_{ij}$ are the tensor perturbations. Working in comoving gauge, one also has $\delta \phi_{\parallel} = 0$ identically and, thus, one is left with two independent scalars, $\zeta$ and $\delta \phi_{\perp}$\,. The quadratic effective action for these scalar fields contains an interaction term between them, whose strength depends on $\eta_{\perp}$\,; see \cite{PSZ,FRPRW} and references therein. Due to that, rapid turns in field space can enhance the amplitude of the curvature perturbation ${\cal P}_{\zeta}$ as follows:
\be
{\cal P}_{\zeta} \,\sim \,{\cal P}_0 \,e^{c \,|\eta_{\perp}|} \,\,\, ,
\ee
where $c = const$ and ${\cal P}_0$ is the CMB level. It is easy to achieve the desired enhancement ${\cal P}_{\zeta}/{\cal P}_0 \sim 10^7$ for actual rapid-turn solutions of the background equations of motion, as shown in \cite{LA}.

\subsection{Dynamical dark energy}

Non-geodesic field space trajectories can also lead to novel effects in multifield models of dark energy. The reason is the following. The main difference between a cosmological constant and dynamical dark energy is that the latter can have fluctuations. So its most important characteristics are: 1) the equation-of-state parameter $w_{\scriptscriptstyle DE}$ and 2) the speed of sound $c_s^{\scriptscriptstyle DE}$ of the dark energy perturbations. In single-field (called quintessence) models, $c_s^{\scriptscriptstyle DE} = 1$ identically.\footnote{We use natural units, in which the speed of light is $c=1$\,.} Hence, observational distinctions from a cosmological constant are difficult to discern, unless the equation of state is significantly different from $w_{\scriptscriptstyle DE} = -1$\,. On the other hand, in the multifield case one can have $c_s^{\scriptscriptstyle DE} < 1$\,. Therefore, new observational features, in the matter distribution inside the cosmological horizon, could occur even when $w_{\scriptscriptstyle DE}$ is close to $-1$; see the discussion in \cite{ASSV}. This makes multifield dark energy particularly interesting phenomenologically. The key reason $c_s^{\scriptscriptstyle DE}$ can be reduced in this class of models is again the existence of background solutions with rapid-turning field-space trajectories, as we will discuss in more detail in the next Section. It is also worth noting that multifield models of dark energy can easily satisfy the various swampland conjectures, as illustrated in \cite{ASSV,EASV,PMB}.

\section{Multifield model of late dark energy}

A particularly interesting model of dynamical dark energy, relying on exact background solutions in the rapid turn regime, was found and studied analytically in \cite{ADGW,ADGW2}. This is a model of the late Universe, which has the potential to alleviate simultaneously the Hubble and $\sigma_8$ tensions. In this section, we review the model and outline the reasons why it is very promising for resolving both of these cosmological tensions.

\subsection{Exact background solutions}

Classes of exact solutions to the background equations of motion (\ref{EoMs})-(\ref{EinstEqs}) can be found by using the hidden symmetry method. The basic idea is to impose the existence of a hidden symmetry on the reduced action, obtained by substituting the Ansatze (\ref{metric_g}) in the action (\ref{Action_gen}). The resulting conditions restrict the form of the scalar potential and, in the two-field case with rotationally-invariant metric $G_{IJ}$, require the scalar field-space to be a hyperbolic surface; for more details, and general solutions of the hidden symmetry conditions, see \cite{ABL1,ABL2,LA3}. Furthermore, by using this symmetry, one can simplify the equations of motion, which greatly facilitates finding exact background solutions (see \cite{ABL1}). 

\begin{figure}[t]
\begin{center}
\includegraphics[scale=0.35]{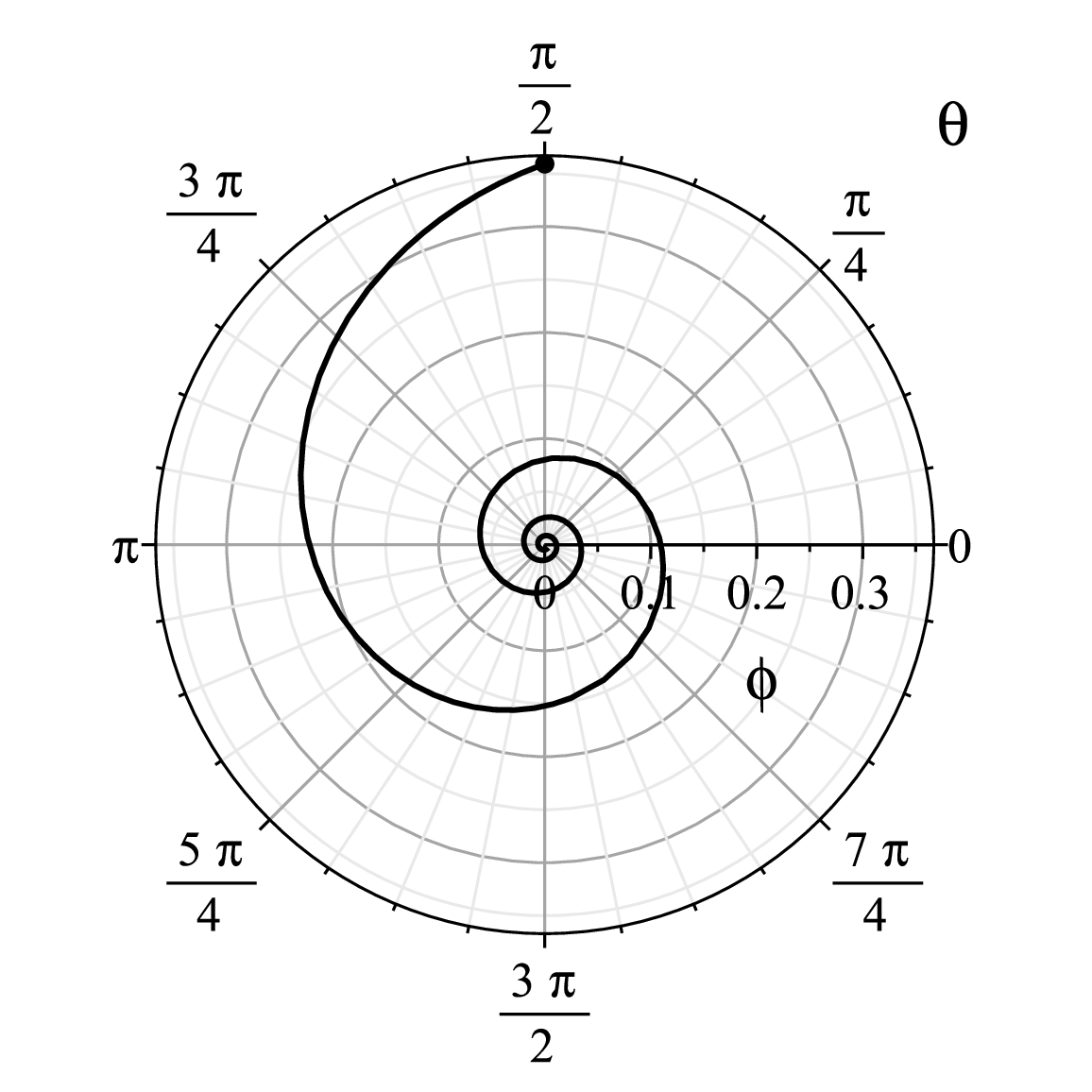}
\hspace*{0.2cm}
\includegraphics[scale=0.35]{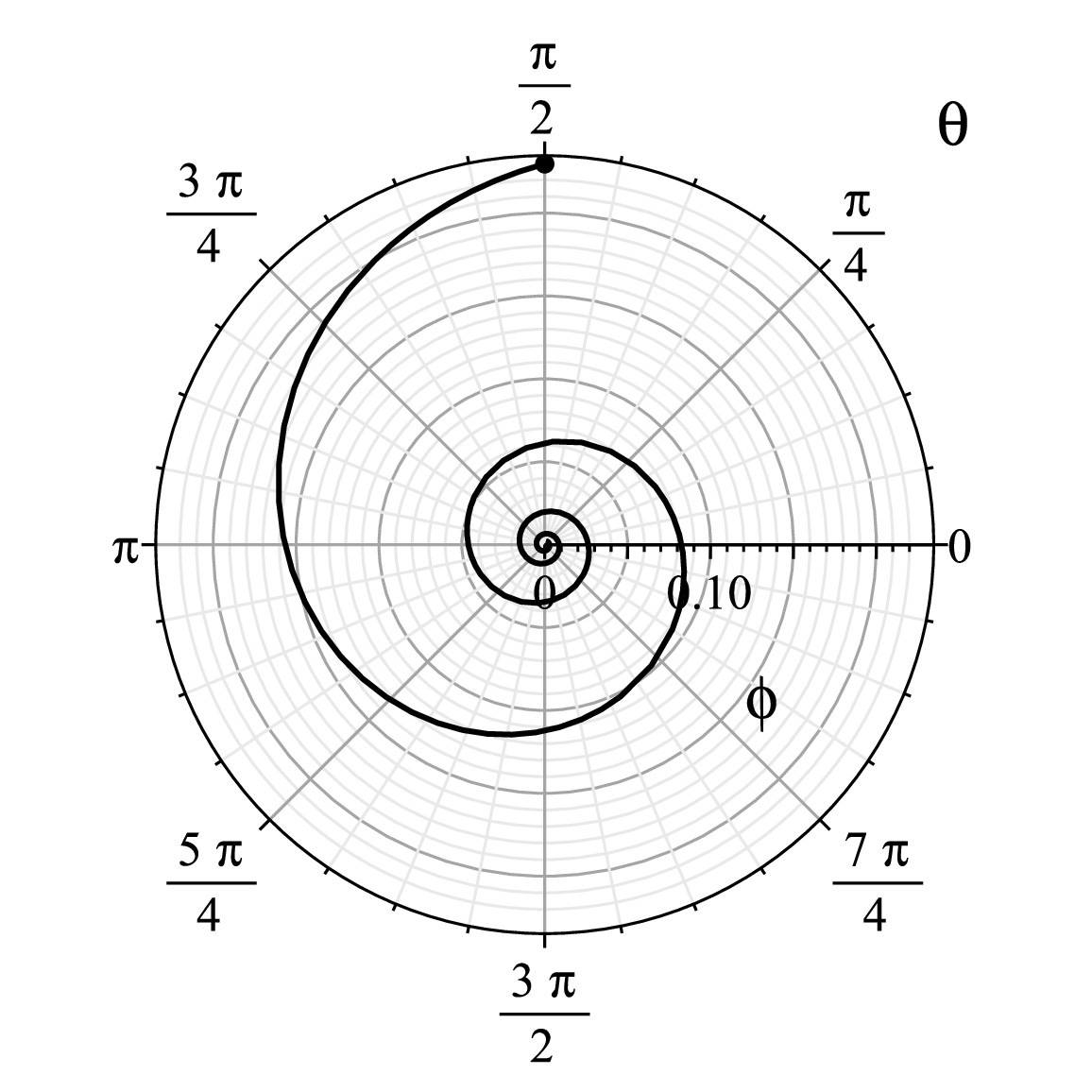}
\end{center}
\vspace{-0.7cm}
\caption{{\small Two examples of field-space trajectories $\left(\varphi(t), \theta(t)\right)$ of the exact solutions of \cite{ADGW}. (The numerical values of the relevant integration constants, used in these plots, can be found in that reference.) The solid dot at $\theta = \frac{\pi}{2}$ and $\varphi \neq 0$ denotes the starting point of the trajectories at an initial moment of time.}}
\label{Traj}
\vspace{0.1cm}
\end{figure}

By adapting this method, \cite{ADGW} found a class of exact solutions describing cosmic acceleration, which are always in the rapid turn regime. These solutions are obtained for specific choices of a field-space metric and a scalar potential. Namely, the scalar field space is taken to be the simplest hyperbolic surface - the Poincar\'e disk. Its metric can be written as:\footnote{A, perhaps, more familiar (in the $\alpha$-attractor and mathematics literature) form of the Poincar\'e-disk metric is the following: $ds^2_G \,= \,6\alpha \,\frac{dz d\bar{z}}{(1-z\bar{z})^2}$\,, where $z = \rho e^{i \theta}$ and $\rho \in [0,1)$\,. This expression can be rewritten as (\ref{Gmetric}) by using the coordinate transformation $\rho \,= \,\tanh \!\left( \frac{\varphi}{\sqrt{6 \alpha}} \right)$ with $\alpha = \frac{16}{9}$\,. Note that this value of the parameter $\alpha$  is singled out by the hidden symmetry used in \cite{ADGW}.}
\be \label{Gmetric}
ds^2_{G} = d\varphi^2 + f(\varphi) d\theta^2 \quad , \quad {\rm where} \quad f (\varphi) \, = \, \frac{8}{3} \,\sinh^2 \!\left( \sqrt{\frac{3}{8}} \,\varphi \right)
\ee
and we have denoted for convenience:
\be \label{Backgr_id}
\phi^1_0 (t) \equiv \varphi (t) \qquad , \qquad \phi^2_0 (t) \equiv \theta (t) \quad .
\ee
In addition, for the scalar potential one takes:
\be \label{Vs}
V (\varphi) \,\, = \,\, C_V \,\cosh^2 \!\left( \frac{\sqrt{6}}{4} \,\varphi \right) - \frac{4}{3} \,\omega^2 \quad , \quad {\rm where} \quad C_V, \omega = const \quad {\rm and} \quad C_V > \frac{4}{3} \omega \quad .
\ee
The functional form of this $V(\varphi)$ is compatible with a certain hidden symmetry for the above system of fields with metric (\ref{Gmetric}). This enables one to simplify significantly the relevant equations of motion, even though the constant term in (\ref{Vs}) breaks that symmetry. The simplified field equations can be easily solved analytically \cite{ADGW}. The resulting exact solutions have trajectories in field space, which continuously spiral toward the center of the Poincar\'e disk; see Figure \ref{Traj} for illustration of the shape of these trajectories. Therefore, this is a class of backgrounds with  strongly non-geodesic field-space trajectories or, in other words, background solutions that are always in the rapid turn regime. Note that, in this class of solutions, $\varphi (t)$ is a monotonically decreasing function and hence the relevant potential, given in (\ref{Vs}), decreases with time. The spacetime of these solutions approaches monotonically de Sitter space, while the equation of state parameter $w_{\scriptscriptstyle DE}$ tends fast to $-1$\,. (For more details, see \cite{ADGW}.) So, as backgrounds describing the late Universe, these solutions are not so different from a cosmological constant. However, the perturbations of the dark energy scalars $\varphi$ and $\theta$ can lead to distinct observational features.

\subsection{Dark energy perturbations}

As in (\ref{Pert_def}), we define the dark energy (DE) perturbations $\delta \phi^I$ around the background as:
\be
\phi^I (t, \vec{x}) = \phi^I_0 (t) + \delta \phi^I (t, \vec{x}) \,\,\, .
\ee
The dynamics of these perturbations, when $(\phi^1_0 (t) , \phi^2_0 (t)) \equiv ( \varphi(t) , \theta(t))$ is the background solution of \cite{ADGW}, was studied in \cite{ADGW2}. Using the rapid turn approximation and neglecting interaction with matter, it was shown there that the sound speed of the DE perturbations is given by the formula:
\be \label{cs}
c_s^{-2} \approx 1 + \frac{4 \Omega^2}{M_T^2 + M_N^2} \,\,\, ,
\ee
where \,$M_T^2 = T^I T^J V_{;IJ}$ \,and \,$M_N^2 = N^I N^J V_{;IJ} - \Omega^2 + \varepsilon H^2 {\cal R}$ \,are the effective masses of the perturbations' components $\delta \phi_{\parallel}$ and $\delta \phi_{\perp}$, respectively, defined as in (\ref{Pert_TNproj}); here we have used the notation $V_{;IJ} \equiv \nabla_I \nabla_J V$ and ${\cal R}$ is the Ricci scalar of the field-space metric $G_{IJ}$\,. From (\ref{cs}) it is evident that rapid turning can lead to a reduced speed of sound. Furthermore, studying this formula, \cite{ADGW2} showed that $c_s^2 (t)$ is a monotonically decreasing function in the class of solutions under consideration. This function reaches fast the lower bound $c_s^2 \approx 0.2$\,, which corresponds to the following sound horizon for the dark energy perturbations:
\be \label{rs}
r_s^{\scriptscriptstyle DE} \approx 6.37\,{\rm Gpc} \,\,\, . 
\ee
The DE perturbations' sound horizon gives a spatial scale, such that on order $r_s^{\scriptscriptstyle DE}$ or larger, dark energy clustering can affect the large-scale structure of the Universe. Interestingly, in the model of \cite{ADGW,ADGW2} the numerical value of this spatial scale, given by (\ref{rs}), is universal in the sense that it does not depend on any of the numerous integration constants in the background solution.

\subsection{Alleviating cosmological tensions}

So far, in the above considerations, matter has been neglected. However, its inclusion is necessary, in order to address a number of cosmological tensions. Recall that, as a function of time, the matter density $\rho_m$ behaves as:
\be \label{rho_m}
\rho_m = \frac{\rho_{m_0}}{a^3 (t)} \qquad , \qquad \rho_{m_0} = const \quad .
\ee
A class of exact solutions, extending the pure DE ones of \cite{ADGW} to the case of non-negligible $\rho_m$\,, was found in \cite{ADGW2}. The scale factor and background scalars in these solutions have the form:
\bea \label{Ch_var}
a (t) &=& \left[ u^2 - \left( v^2 + w^2 \right) \right]^{1/3} \,\,\,\, , \nn \\
\varphi (t) &=& \sqrt{\frac{8}{3}} \,{\rm arccoth} \!\left( \sqrt{\frac{u^2}{v^2 + w^2}} \,\,\right) \,\,\,\, , \nn \\
\theta (t) &=& \theta_0 + {\rm arccot} \!\left( \frac{v}{w} \right) \,\,\,\, , \,\,\,\, \theta_0 = const \quad ,
\eea
where $u$, $v$ and $w$ are the following functions: 
\bea \label{Sols_uvw}
u (t) &=& C_1^u \sinh (\tilde{\kappa} t) + C_0^u \cosh (\tilde{\kappa} t) \quad , \nn \\
v (t) &=& C_1^v \sin (\omega t) + C_0^v \cos(\omega t) \quad , \nn \\
w (t) &=& C_1^w \sin (\omega t) + C_0^w \cos(\omega t) \quad .
\eea
Here $C_{0,1}^{u,v,w} = const$ and
\be \label{tkappa}
\tilde{\kappa} \equiv \frac{1}{2} \sqrt{3 C_V - 4 \omega^2} \,\,\, .
\ee
In addition, the constants in (\ref{rho_m}) and (\ref{Sols_uvw})-(\ref{tkappa}) are related by:
\be \label{Constr}
\tilde{\kappa}^2 \!\left[ (C_0^u)^2 - (C_1^u)^2 \right] + \omega^2 \left[ (C_0^v)^2 + (C_1^v)^2 + (C_0^w)^2 + (C_1^w)^2 \right] + \frac{3 \rho_{m_0}}{4} = 0 \,\,\, .
\ee
In other words, there is one less independent integration constant than is apparent in (\ref{Sols_uvw}). 

One can easily verify that the backgrounds (\ref{Ch_var})-(\ref{Constr}) solve exactly the relevant equations of motion, namely:
\be \label{EinstEq_m}
3H^2 = \frac{1}{2} \!\left( \dot{\varphi}^2 + f \dot{\theta}^2 \right) + V + \rho_m \quad ,
\ee
\be \label{ScalarEoMs}
\ddot{\varphi} - \frac{f'}{2} \dot{\theta}^2 + 3 H \dot{\varphi} + \pd_{\varphi} V = 0 \qquad , \qquad \ddot{\theta} + \frac{f'}{f} \dot{\varphi} \dot{\theta} + 3 H \dot{\theta} + \frac{1}{f} \pd_{\theta} V = 0  \quad ,
\ee
when $f$ and $V$ are given by (\ref{Gmetric}) and (\ref{Vs}) respectively. Also, the pure dark energy solutions of \cite{ADGW} are obtained by taking \,$C_1^v,C_0^w = 0$ \,, \,$C_0^v = C_1^w \equiv C_w$ and $\rho_{m_0}=0$ \,inside (\ref{Sols_uvw}) and (\ref{Constr}). In this special case, one has $\dot{\theta} = \omega$\,. However, the more general class of solutions (\ref{Ch_var})-(\ref{Constr}) allows for a varying $\dot{\theta}$\,.

As explained in \cite{ADGW2}, the above class of exact solutions is suitable for describing the cosmology of the late Universe, and specifically the transition from matter domination to the current DE epoch. For that purpose, one needs to take a part of parameter space, in which at some initial moment the dark energy density is negligible compared to the matter density. This can be achieved by choosing suitably the (so far) arbitrary overall scale of the DE contribution. Then, one would have as an initial condition the relation $3 H^2 \approx \rho_m$\,, which is standard during matter domination. During that epoch, the functions $f$ and $V$ will remain almost constant, since the DE scalars $\varphi$ and $\theta$ are effectively frozen due to the large Hubble friction. As $\rho_m$ decreases with time, at some point it will become comparable to the energy density of the DE fields. That is when $\varphi$ will start rolling toward the minimum of $V$, situated at $\varphi = 0$ according to (\ref{Vs}). So $V$ will start decreasing and will tend to a constant again, at late times.

In \cite{ADGW2} it was argued that this multifield cosmological model is suitable for alleviating simultaneously the Hubble and $\sigma_8$ tensions. The Hubble tension is an apparent contradiction between the value of the Hubble constant $H_0$, extracted from early Universe data\footnote{More precisely, this is the value of the Hubble parameter in the early Universe inferred from the CMB, which is evolved to the present with the standard $\Lambda$CDM cosmological model.}, and its value deduced from observations of local objects (like supernovae) in the late Universe. Notably, the increasing precision of the observational data on both sides, i.e. the CMB and the late Universe, has led to an increase in the disagreement between the two $H_0$ values. One can resolve this tension in a cosmological model, which leads to a greater value of the Hubble constant, compared to the prediction of $\Lambda$CDM, when starting from the same value (as in $\Lambda$CDM) of the Hubble parameter $H(t)$ at some initial time, for instance during matter domination. The dynamical DE model discussed above allows for precisely this kind of behavior in a part of its parameter space, as argued in \cite{ADGW2}. Even more interestingly, our model seems suitable for alleviating the $\sigma_8$ tension everywhere in its parameter space. The quantity $\sigma_8$ measures the amplitude of the linear matter perturbations on a scale of order 10 Mpc. Inferring its magnitude from early Universe data, evolved to today via $\Lambda$CDM, leads to a discrepancy with the $\sigma_8$ magnitude determined from late Universe observations, similarly to the Hubble tension (albeit to a lesser degree). In the model of \cite{ADGW,ADGW2}, on the other hand, one always expects a lower value of $\sigma_8$ today, compared to $\Lambda$CDM, thus alleviating (or even resolving) the tension. The key to this conclusion is in the specific behavior of $w_{\scriptscriptstyle DE}$ and $c_s^{\scriptscriptstyle DE}$ in our dynamical DE model. Although these quantities are time-dependent in our case, they tend fast to the bounds $w_{\scriptscriptstyle DE} \approx -1$ and $c_s^{\scriptscriptstyle DE} \approx 0.447$, for any values of the integration constants in our solutions \cite{ADGW2}. These numerical values correspond to ${\sigma}_8^{\rm {\scriptscriptstyle DE}} < {\sigma}_8^{\rm {\scriptscriptstyle \Lambda \!CDM}}$, according to the results of \cite{MT} which investigated the general dependence of $\sigma_8$ on the DE equation-of-state parameter and DE perturbations' sound speed. Thus, our model should alleviate the $\sigma_8$ tension everywhere in its parameter space, although the precise magnitude of $\sigma_8$ will, of course, depend on the concrete choices of values for the integration constants. 

Much more involved (and, likely, numerical) computations are required, in order to establish what parts of the parameter space of the model can be (most) compatible with all current observations. We hope to report on this problem in the near future.

\section{Concluding remarks}

Dynamical dark energy models have been of theoretical interest for a while, in the search for alternatives to $\Lambda$CDM that can resolve various cosmological tensions and/or explain aspects of the dark Universe. In addition, a great observational motivation to study them has emerged very recently. Namely, the new release of data from the DESI collaboration indicates an evolving DE equation of state \cite{DESI}. The statistical significance of this result is not yet above the discovery threshold. Nevertheless, the DESI data already present a serious challenge to the standard cosmological model. Hence, it is very important to understand whether the conclusions of \cite{DESI} could change qualitatively, if one were to use a different parametrization of $w_{\scriptscriptstyle DE}$ than the phenomenological one adapted there. This question was addressed in \cite{CCSGVM}, where a number of dynamical models were investigated and compared to the observational data from DESI, as well as to several supernovae datasets. The analysis in that work confirms substantial evidence for evolving dark energy, while inferring a preference for $w_{\scriptscriptstyle DE} < -1$. The last inequality is very problematic theoretically, as it implies a violation of the null energy condition. See, however, \cite{BOVW} for a string-inspired model that can lead to such a phantom behavior effectively, without relying on exotic forms of energy. Furthermore, it was shown in \cite{SP} that a phantom DE equation of state can be an artifact of the phenomenological parametrization $w_{\scriptscriptstyle DE} (a) = w_0 + w_a \left( 1-a/a_0 \right)$\,, which is widely used in the literature (including in \cite{DESI}) for analyzing the observational data; here $a_0$ is the present-day scale factor, often set to $a_0 = 1$\,, and $w_{0,a} = const$\,.\footnote{In line with these considerations, \cite{AACHLMS} proposed recently a particular two-field model, which does not violate the null energy condition, but is nevertheless consistent with the DESI data.}

It would be very interesting to perform the kind of analyses, done in \cite{CCSGVM} and \cite{SP}, for the multifield model of \cite{ADGW,ADGW2}. The goal would be to explore whether the peculiar novel features of the rapid turn regime could lead to agreement with the observational data, without the need for a phantom-like behavior. If not, then it would be worth studying whether the inclusion of an interaction with dark matter (of a particular type, in the vein of \cite{BOVW}) could improve the model phenomenologically.\footnote{It is worth commenting on claims in the literature that quintessence, and more generally models with $w (z) \ge - 1$ for any $z$\,, can only worsen the Hubble tension; see for instance \cite{CPSJY}. Such arguments either use particular quintessence models or, as in \cite{CPSJY}, rely on tacit assumptions that need not be true. Of course, in models with $w(z) > -1$ the energy density of dark energy, $\rho_{\scriptscriptstyle de} (z)$\,, increases with $z$\,. However, the matter density $\rho_{\scriptscriptstyle m} (z)$ has to increase much faster, to ensure that $\rho_{\scriptscriptstyle m} \gg \rho_{\scriptscriptstyle de}$ and thus there is matter domination in the appropriate redshift range. So the evolution of the Hubble parameter is affected by $\rho_{\scriptscriptstyle de}$ only at small $z$\,. And one can obtain a somewhat greater (than in $\Lambda$CDM) Hubble constant $H_0$\,, thereby alleviating the Hubble tension, {\it together} with a somewhat lower (than in $\Lambda$CDM) present-day matter-density fraction $\Omega_m$\,, which is allowed by current observations. This is, indeed, what happens in a suitable part of the parameter space of the model of \cite{ADGW,ADGW2}, as will be shown in a forthcoming publication.} 

\section*{Acknowledgements}
I have received partial support from the Bulgarian NSF grant KP-06-N88/1. I am also grateful to the Simons Physics Summer Workshop at Stony Brook University for hospitality during the completion of this work.

\end{document}